\documentclass[aps,showpacs,superscriptaddress,twocolumn]{revtex4}%
\usepackage{amsfonts}
\usepackage{mathbbold}
\usepackage{amsmath}
\usepackage{amssymb}
\usepackage{graphicx}%
\begin{document}
\title{ Oriented gap opening in the
magnetically ordered state of Iron-pnicitides: an impact of intrinsic  unit cell doubling on  the $Fe$ square lattice by $As$ atoms}
\author{Ningning Hao}
\affiliation{Beijing National Laboratory for Condensed Matter Physics and Institute of
Physics, Chinese Academy of Sciences, P. O. Box 603, Beijing 100190, China}
\affiliation{Department of Physics, Purdue University, West Lafayette, Indiana 47907, USA}
\author{Yupeng Wang}
\affiliation{Beijing National Laboratory for Condensed Matter Physics and Institute of
Physics, Chinese Academy of Sciences, P. O. Box 603, Beijing 100190, China}
\author{Jiangping Hu}
\affiliation{Beijing National Laboratory for Condensed Matter Physics and Institute of
Physics, Chinese Academy of Sciences, P. O. Box 603, Beijing 100190, China}
\affiliation{Department of Physics, Purdue University, West Lafayette, Indiana 47907, USA}

\begin{abstract}
We show that the complicated band reconstruction near Fermi surfaces in the
magnetically ordered state of iron-pnictides observed by angle-resolved
photoemission spectroscopies (ARPES)  can be  understood in a
meanfield level if the \emph{intrinsic unit cell doubling} due to As atoms is
properly considered as shown in the recently constructed S$_{4}$ microscopic
effective model. The (0,$\pi$) or ($\pi$,0) col-linear antiferromagnetic
(C-AFM) order does not open gaps between two points at Fermi
surfaces linked by the ordered wave vector but 
forces a band reconstruction involving four points in unfolded Brillouin zone
(BZ) and gives rise to small pockets or hot spots.  The S$_4$ symmetry  naturally chooses   a staggered orbital order  over a ferro-orbital order to coexist with the C-AFM order.   These results strongly suggest that  the
kinematics based on the S$_{4}$ symmetry captures the essential low energy
physics of iron-based superconductors.
\end{abstract}

\pacs{xxxx}
\maketitle

\emph{Introduction:} The parent compounds of all Iron-pnictide high
temperature superconductor\cite{Hosono,ChenXH,wangnl2008} are characterized by a unique  C-AFM order\cite{spinwave-dela2008, zhao1, DaiHureview}. In the past several years,
although the majority of theoretical studies on these compounds have
successfully obtained such a magnetic state from both itinerant and local
moment models\cite{Dongj2008,Si2008,Fang2008nematic,Hu2012u,mahu2012,Ma2008lu,wangfa2009,Kuroki2011}, the band reconstruction in the C-AFM state observed by ARPES\cite{miyi2009,richardarpes2010,nematic-yi2010,zhangapres2012}
remains a puzzle. Unlike in a conventional spin density wave (SDW)  state where gaps at any two points at Fermi
surfaces which are connected by the ordered SDW wave vector should be opened
generically, the band structure in the C-AFM state appears to be reconstructed by the development
of the magnetic order. While optical measurements suggest a significant portion of Fermi surfaces are gapped\cite{huoptical2009,huoptical2009b,optical-chen2010,Moonoptical2010},  ARPES observes that
small pockets or hot spots develop around Fermi surfaces. It is fair to
say that the scenario of the gap opening due to a magnetic order in a
multi-orbital system, such as iron-pnictides, is complex. For example, it has been argued that the entire Fermi surfaces can be gaped out\cite{ranying2009}.  Nevertheless, the surprising band
reconstruction is never clearly understood until now.

\begin{figure}[ptb]
\begin{center}
\includegraphics[width=1.0\linewidth]{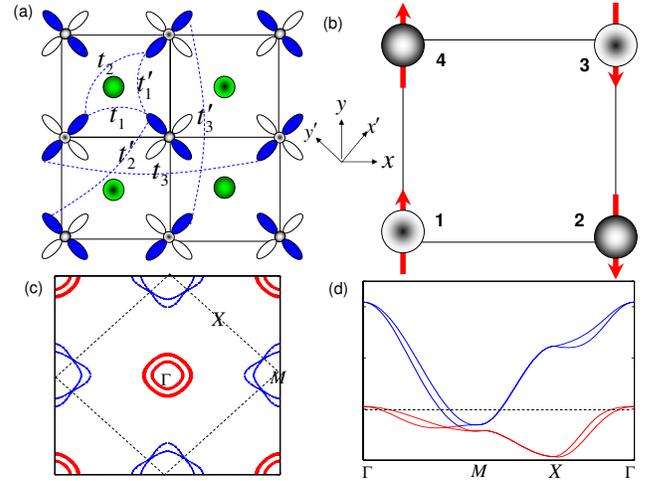}
\end{center}
\caption{(color online) (a) The lattice structure of Fe-As trilayers. The
different filled green/big balls indicate the top and bottom As layer. The
bule/deep and white/light orbital patterns indicate two S$_{4}$ iso-spin
components. The effective hopping parameters up to 3rd nearest neighbor are
marked. (b) The ($\pi,0$) C-AFM order are shown. Four Fe sublattices in the
unit cell of C-AFM state are marked. (c) and (d) The Fermi surfaces and energy
spectrums along high symmetric lines for normal paramagnetic state are shown.
The hopping parameters are set $t_{1}=0.47$, $t_{1}^{\prime}=0.53$,
$t_{2}=0.9$, $t_{2}^{\prime}=-0.18$, $t_{3}=0.0$, $t_{3}^{\prime}=0.1$,
$t_{c}=0.1$ and $\mu=-0.5$ for electron underdoped 4.5\%.}\label{fig1}
\end{figure}

In this paper, we demonstrate that in the recently constructed effective
two-orbital model with the S$_{4}$ symmetry, which is the symmetry of the
trilayer FeAs structure, the band reconstruction becomes a natural consequence
from the intrinsic unit cell doubling due to As atoms\cite{Hu2012s4}. 
The spirit of
the S$_{4}$ model is that the kinematics can be divided into weakly coupled
two orbital models controlled separately by two S$_{4}$ iso-spin components as shown in Fig.\ref{fig1}(a). For each S$_{4}$ iso-spin component, a unit cell includes two
irons, namely, the unit cell is doubled, and the doubling results in the formation
of a pair of bands which are responsible for a hole pocket at $\Gamma$ and an
electron pocket at $M$. Although this kinematics is hidden in
various multi-orbital models constructed for iron-based superconductors as
shown in\cite{Hu2012s4}. Here we show that it is the extraction of such a minimum
effective model allows us to naturally capture the band reconstruction
features mentioned above in a simple mean-field Hamiltonian. The S$_4$ symmetry also  naturally  imposes   a staggered orbital order  over a ferro-orbital order to coexist with the C-AFM order. \emph{ Our results  suggest that  a proper starting point in kinematics is crucial to obtain a correct meanfield result and  the kinematics
based on S$_{4}$ symmetry due to As atoms provides the essential low energy physics of
iron-based superconductors.}

\begin{figure}[ptb]
\begin{center}
\includegraphics[width=1.0\linewidth]{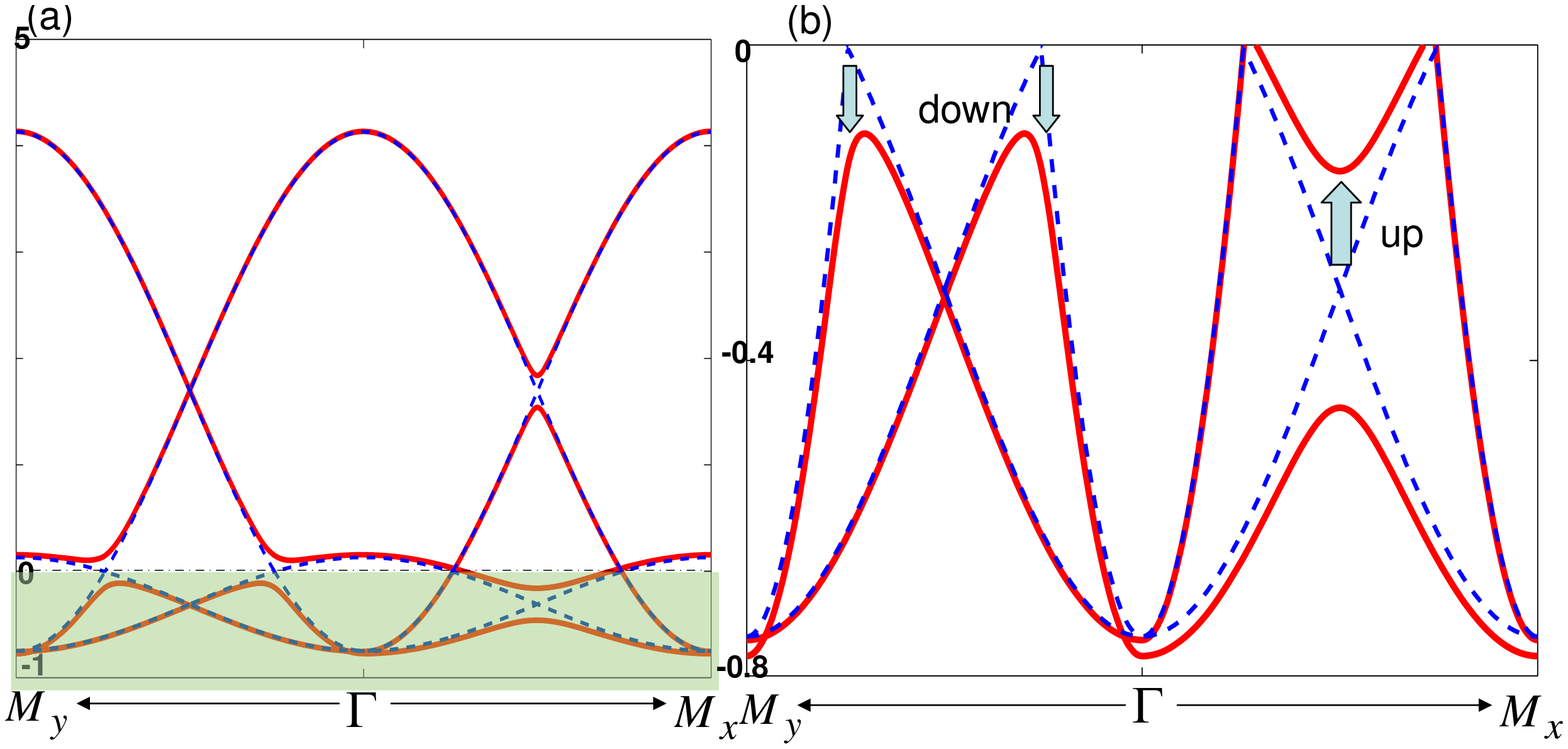}
\end{center}
\caption{(color online) (a) The energy bands of Hamiltonian Eq.
(\ref{H_single band}) with spin-up component in ($\pi,0$) C-AFM state along
$\Gamma-M_{x}$ and $\Gamma-M_{y}$ directions are shown. The bule/dashed lines
indicate the normal state while the red/solid lines indicate the C-AFM state.
The shadow regime indicates the parts below Fermi surfaces. (b) The shadow
regime in (a) is zoomed in. The arrows indicate the bands in C-AFM state shift
up or down compared with that in normal state. We set $t_{1}=t_{1}^{\prime
}=0.5$, $t_{2}=0.9$, $t_{2}^{\prime}=-0.18$, $t_{3}=t_{3}^{\prime}=0.0$,
$\mu=-0.69$ and $\Delta_{SDW}=0.15$.}%
\end{figure}

\emph{ Oriented gap opening mechanism with unit cell doubling: } The  effect  of the unit cell doubling   on  the C-AFM state  can be elucidated  by considering   the kinematics of a  single S$_4$ iso-spin component. Deeply in a C-AFM state, regardless of the origin of the magnetic ordering, we should expect that a SDW order  can  capture the basic features of  itinerant bands.   As shown in Fig.\ref{fig1}(a),  the kinematics of the  single   blue (deep) orbital  has an intrinsic unit cell doubling due to the large difference between the two next nearest neighbor hopping parameters, $t_2$ and $t_2'$ (note: the details of the S$_{4}$ model can be found in our previous paper\cite{Hu2012s4}). 

In the $Q_1=(\pi,0)$ C-AFM state shown in Fig. 1 (b), at the 
1 and 4 sites,  the  spin-up component is a majority while at the sites 2 and 3, the   spin-down component is a majority.  A simple SDW order is  thus given by 
$\Delta_{SDW}=<\sum_{k\sigma}\eta_\sigma c_{k,\sigma}^{\dag}c_{k+Q_1,\sigma}>$ where  $\eta_{\uparrow}=1$ and $\eta_{\uparrow}=-1$ for different spin
components. Due to the intrinsic unit cell doubling, for any given point $k$ in unfolded BZ zone, there are four momentums linked in the presence of the magnetic order, $k, k+Q_1,k+Q$, and $ k+Q_2$ where  $Q=(\pi,\pi)$ and $Q_{2}=(0,\pi)$.   The mean-field  Hamiltonian that described the $(\pi,0)$ C-AFM state can thus be expressed as
$H=\sum_{k,\sigma}\psi_{k,\sigma}^{\dag}\mathcal{H}(k,\sigma)\psi_{k,\sigma}$
with $\psi_{k,\sigma}=[c_{k,\sigma},c_{k+Q,\sigma},c_{k+Q_{1},\sigma
},c_{k+Q_{2},\sigma}]^{T}$. $\mathcal{H}(k,\sigma)$ is given by \emph{ }\ \ \
\begin{equation}
\mathcal{H}(k,\sigma)=\left[
\begin{array}
[c]{cccc}%
f(k) & g(k) & \eta_{\sigma}\Delta_{SDW} & 0\\
g^{\ast}(k) & f(k+Q) & 0 & \eta_{\sigma}\Delta_{SDW}\\
\eta_{\sigma}\Delta_{SDW} & 0 & f(k+Q_{1}) & g(k+Q_{1})\\
0 & \eta_{\sigma}\Delta_{SDW} & g^{\ast}(k+Q_{1}) & f(k+Q_{2})
\end{array}
\right]  \label{H_single band}%
\end{equation}
where $f(k)=2t_{1s}(\cos k_{x}$+$\cos k_{y})$+$2t_{1d}(\cos k_{x}-\cos k_{y})$
$+4t_{2s}\cos k_{x}\cos k_{y}+2t_{3s}(\cos2k_{x}+\cos2k_{y})$
+$2t_{3d}(\cos2k_{x}-\cos2k_{y})-\mu$, $g(k)$=$4t_{2d}\sin k_{x}\sin k_{y}$
and $t_{is}=(t_{i}+t_{i}^{\prime})/2$, $t_{id}=(t_{2}-t_{2}^{\prime})/2$ with
$i=1,2,3$.  It is explicit to find that the spin-up and spin-down electrons
play a symmetrical role in building up the $(\pi,0)$ C-AFM states. Hence, we
can only investigate $\mathcal{H}(k,\uparrow)$.  

Now we are in a position to show that the gap opening described in Eq.\ref{H_single band} is rather different from a conventional one in a conventional SDW  state. Here the energy dispersions of $\mathcal{H}(k,\uparrow)$ can not be given by a simple  analytical form.  We provide the numerical results for spin-up case shown in Fig. 2.
It is easy to see that  the band reconstruction in the C-AFM state are
anisotropic along the $\Gamma-M_{x}$ and $\Gamma-M_{y}$ directions. Below
Fermi surfaces, an energy splitting  takes place between two hole bands at ($\frac{\pi}{2},0$)
point along  the $\Gamma-M_{x}$.  Along the $\Gamma-M_{y}$ direction, gaps at Fermi surfaces are opened between a hole band and an electron band.  Due to the fact that the gap openings are different along different directions, we call this behavior as \emph{oriented gap opening}.  As we will show later, this behavior leads to a development of hot spots or a small pockets on Fermi surfaces.  

\emph{The self-consistent mean-field results:}
In the above analysis, the magnetic order is introduced artificially. To show the magnetic order is generated by standard interactions in the S$_4$ model (note, a recent quantum monte carlo calculation has already shown the C-AFM correlation is the leading magnetic correlation of the model near half filling\cite{mahu2012})  as well as  to produce   more realistic results that can be closely compared to ARPES observations,   we apply a self-consistent meanfield calculation to consider the full S$_{4}$ model with the full interaction terms. 

Including all possible electron-electron interaction terms, the general  Hamiltonian of the S$_4$ model is given by\cite{Hu2012s4},%
\begin{equation}
H=H_{0}+H_{int} \label{H_tot}%
\end{equation}
where $H_0$ is the kinematics part  given in\cite{Hu2012s4} and %
\begin{align}
H_{int}  &  =U\sum_{i,\alpha}(\hat{n}_{i\alpha\uparrow}^{c}\hat{n}%
_{i\alpha\downarrow}^{c}+\hat{n}_{i\alpha\uparrow}^{d}\hat{n}_{i\alpha
\downarrow}^{d})+U^{\prime}\sum_{i,\alpha,\sigma}\hat{n}_{i\alpha\sigma}%
^{c}\hat{n}_{i\alpha\bar{\sigma}}^{d}\nonumber\\
&  +(U^{\prime}-J_{H})\sum_{i,\alpha,\sigma}\hat{n}_{i\alpha\sigma}^{c}\hat
{n}_{i\alpha\sigma}^{d}\nonumber\\
&  +J_{H}\sum_{i,\alpha}(c_{i\alpha\uparrow}^{\dag}d_{i\alpha\downarrow}%
^{\dag}c_{i\alpha\downarrow}d_{i\alpha\uparrow}+c_{i\alpha\uparrow}^{\dag
}c_{i\alpha\downarrow}^{\dag}d_{i\alpha\downarrow}d_{i\alpha\uparrow}+h.c.)
\label{H_int}%
\end{align}
Here, we use $c$ and $d$ to identify the two S$_{4}$ iso-spin components.  $\hat
{n}_{\alpha i\sigma}^{c}=c_{\alpha i\sigma}^{\dag}c_{\alpha i\sigma}$ and
$\hat{n}_{\alpha i\sigma}^{d}=$ $d_{\alpha i\sigma}^{\dag}d_{\alpha i\sigma}$.
$\alpha$ labels the two sublattices of the iron square lattice with $\alpha=1,2$.  $U$ and $U^{\prime}$ are the
direct intra- and inter-orbital Coulomb repulsions, respectively. $J_{H}$ is
the Hund coupling satisfying $U^{\prime}=U-2J_{H}$. The last term in $H_{int}$
describes the spin-flip term of intro-orbital exchange and  the pair hopping.
Here we neglect the pair hopping term.\begin{figure}[pb]
\begin{center}
\includegraphics[width=1.0\linewidth]{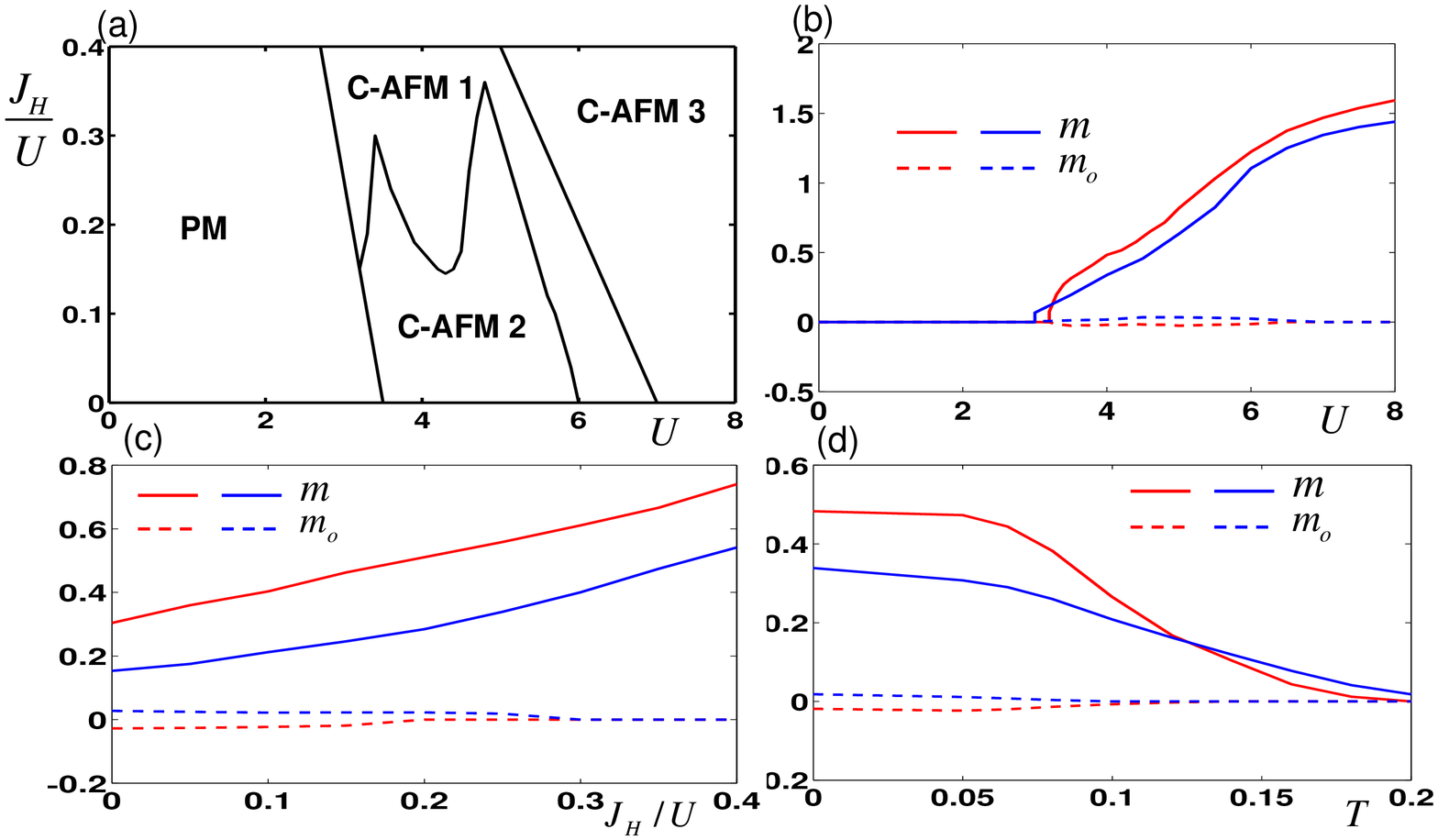}
\end{center}
\caption{(color online) (a) The zero-temperature mean-field phase diagram of
the S$_{4}$ effective model in the $J_{H}/U$ vs $U$ plane in undoped case. We denote 
`PM' for  the paramagnetic phase, `C-AFM 1'  for  the metallic C-AFM phase,
`C-AFM 2' for  the metallic C-AFM phase with a staggered orbital order and  `C-AFM 3' for the
insulating C-AFM phase. (b), (c) and (d) show the C-AFM magnetization
$m$ and stagger orbital polarization $m_{o}$ as the function of $U$, $J_{H}/U$
and $T$. The red color denotes the undoped case while the blue color denotes the 4.5\%
electron-doped case. The parameters  are   $J_{H}/U=0.15$ for undoped case and $0.25$ for 4.5\%
electron doped case in (b),   $U=4$ for both cases in (c), and  $J_{H}/U$ with the  same values as (b)
 and $U=4$ in (d). The hoping parameters are indicated in the caption of
Fig. 1.}%
\end{figure}

To obtain the possible ground-state  in the undoped and underdoped cases, we apply a mean-field
decoupling to the model in  Eq.(\ref{H_int}). 
$H_{int}$ can be simply decoupled by defining the mean-field values of
the diagonal operators,%

\begin{eqnarray}
& & <c_{\alpha\sigma}^{\dag}c_{\alpha\sigma}>=n_{\alpha\sigma}^{c}\text{,
}<d_{\alpha\sigma}^{\dag}d_{\alpha\sigma}>=n_{\alpha\sigma}^{d} \label{n_c}%
\\
& & <c_{\alpha\uparrow}^{\dag}c_{\alpha\downarrow}>=\kappa_{\alpha}^{c}%
,\ <d_{\alpha\uparrow}^{\dag}d_{\alpha\downarrow}>=\kappa_{\alpha}^{d}
\label{kappa_c}%
\end{eqnarray}
If we  assume that  possible ordered states are only allowed to break translation symmetry up to  four irons and let $\alpha=1,2,3,4$ label sites in the four different sublattices respectively as shown in Fig.\ref{fig1}(b),
the mean-field Hamiltonian of Eq.(\ref{H_tot}) in the reciprocal space can be
written as%
\begin{align}
H_{mf}  &  =H_{0}+\sum\limits_{k,\alpha,\sigma}[Un_{\alpha\bar{\sigma}%
}^{c}+U^{\prime}n_{\alpha\bar{\sigma}}^{d}+(U^{\prime}-J_{H})n_{\alpha\sigma
}^{d}]\hat{n}_{k\alpha\sigma}^{c}\nonumber\\
&  +\sum\limits_{k,\alpha,\sigma}[Un_{\alpha\bar{\sigma}}^{d}+U^{\prime
}n_{\alpha\bar{\sigma}}^{c}+(U^{\prime}-J_{H})n_{\alpha\sigma}^{c}]\hat
{n}_{k\alpha\sigma}^{d}\nonumber\\
&  -J_{H}\sum_{k,\alpha}\left[  \kappa_{\alpha}^{c}d_{k\alpha\downarrow}%
^{\dag}d_{k\alpha\uparrow}+(\kappa_{\alpha}^{d})^{\ast}c_{k\alpha\uparrow
}^{\dag}c_{k\alpha\downarrow}+h.c.\right]  +C \label{H_mf}%
\end{align}
where $C$  represents non-operator terms.

The above mean-field Hamiltonian can be numerically solved with a standard
self-consistent procedure by minimizing  free energy. During the iterative
procedure of the self-consistent calculation, we enforced the average electron number per site  at each step remains a constant, namely, keeps the density constant. We define the magnetization
$m_{\alpha}$ and polarization $\Delta n_{\alpha}$ as%
\begin{align}
m_{\alpha}  &  =(n_{\alpha\uparrow}^{c}-n_{\alpha\downarrow}^{c}%
)+(n_{\alpha\uparrow}^{d}-n_{\alpha\downarrow}^{d})\label{m_a}\\
\Delta n_{\alpha}  &  =(n_{\alpha\uparrow}^{c}+n_{\alpha\downarrow}%
^{c})-(n_{\alpha\uparrow}^{d}+n_{\alpha\downarrow}^{d}) \label{n_d}%
\end{align}
 All different  kinds of  standard magnetic orders (such as C-AFM order, ferromagnetic order
and antiferromagnetic order) and orbital orders (such as ferro-orbital order,
staggered-orbital order  and stripe orbital order) are included by above meanfield ansatz.

The zero-temperature phase diagram for the undoped case was obtained and shown
in Fig. 3 (a). The C-AFM order was found to be stable in a broad and
realistic range of interaction strength. Furthermore,   the C-AFM orders exhibit differently  depending on  $U$ and $J_{H}/U$.   When $J_{H}/U$ is large in a moderate
value  of $U$, no orbital-order was  found.  When $J_{H}/U$ is small enough, a small staggered-orbital order coexists with the magnetic order.    When  $U$ is large enough, as expected, the metallic C-AFM order becomes insulating.  It is important to note that for a coexisting state with both C-AFM
order and staggered orbital order, the corresponding magnetizations $m_{\alpha}$ at the sites labeled in Fig.\ref{fig1}(b) are  $m_{1/4}=m$ and $m_{2/3}=-m$  and  the orbital orders are given by $\Delta
n_{i}=m_{o}$ ($i=1,2,3,4$). In Fig. 3 (b) (c) and (d), we plotted $m$ and $m_{o}$
as functions of $U$, $J_{H}/U$ and temperature $T$. From Fig. 3 (b), for the two
undoped and 4.5\% electron doped cases studied here, $m$ jumps into a finite value at
$U\sim3.2$ and $\sim3$ respectively, which means the transition from the PM to the C-AFM is a
first order phase transition. From Fig. 3(c),
the large $J_{H}/U$ enhances the  C-AFM order $m$ and suppresses the staggered orbital order $m_{o}$
in both undoped and underdoped cases. From Fig. 3 (d),  increasing
temperature  suppresses both C-AFM order $m$ and  the staggered orbital order $m_{o}$, and so does the doping.
We also find that increasing doping can
continually reduce the C-AFM order $m$ at low temperature.  All the results are  qualitatively consistent with
the experimental observations\cite{johnstonreview}. 

 The result that a staggered orbital order is favored over a ferro-orbital order  stems from   the S$_4$ symmetry. As shown in Fig.\ref{fig1}(a),  each S$_4$ iso-spin component is formed by two different types of orbitals in the two different sublattices of  the iron square lattice. A staggered orbital order corresponds to the S$_4$ symmetry breaking.    Many theoretical proposals based on models with one-iron per unit cell  emphasize the ferro-orbital order\cite{orbital-saito2010,lv2010-orbital,kruger2008}.   Therefore, in  principle,  the result can be used as a test for  the S$_4$ symmetry. 
Unfortunately,  the staggered orbital order is rather small in our meanfield results.  

  \begin{figure}[ptb]
\begin{center}
\includegraphics[width=1.0\linewidth]{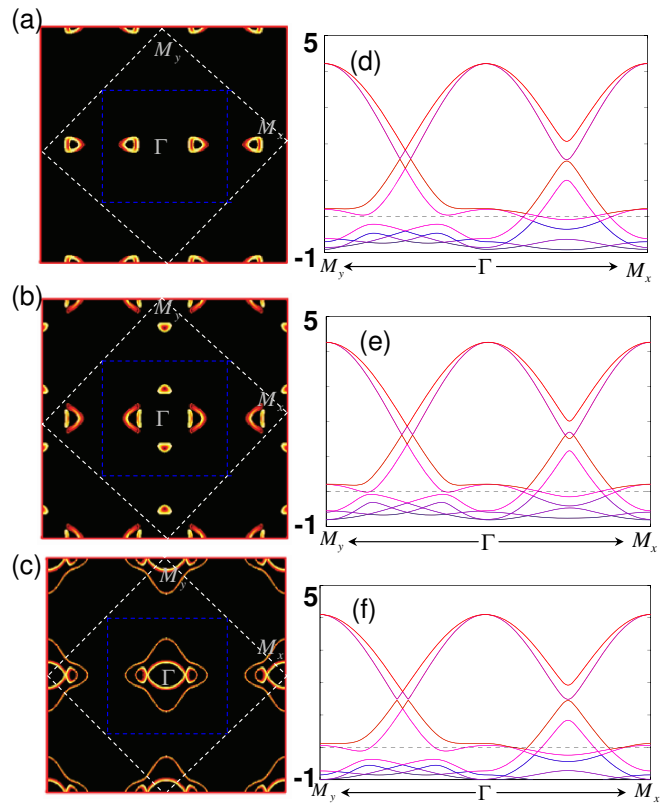}
\end{center}
\caption{(color online) The Fermi surfaces in C-AFM states are shown for some
different interaction strength and doping. (a) the undoped case with $U=4$
and\ $J_{H}/U=0.15$. (b) the undoped case with $U=3.5$ and\ $J_{H}/U=0.15$.
(c) the 4.5\% electron-doped case with $U=4$ and $J_{H}/U=0.25$. (d) (e) and
(f) are the energy bands along $\Gamma-M_{x}$ and $\Gamma-M_{y}$ directions,
respectively.}%
\end{figure}

\emph{The formation of hot spots or small pockets in the C-AFM state:} To obtain further insights into the characters of the C-AFM ordered states, we   investigated the Fermi surface construction which can be
directly observed by APRES experiments.  The
reconstructed Fermi surfaces and bands of our model 
 with  different interaction strengths and doping cases are shown in Fig. 4. The interaction strength
is chosen in the moderate strength range since iron-pnicitides are most likely in intermediate coupling region.   The Fermi
surfaces in Fig. 4 (a), (b) and (c) show a close match to ARPES results measured in
BaFe$_{2}$As$_{2}$\cite{richardarpes2010}, NaFeAs\cite{nematic-yi2010,zhangapres2012} and Ba(Fe$_{1-x}%
$Co$_{x}$)$_{2}$As$_{2}$\cite{miyi2009}, respectively. Furthermore, the  magnetization value in Fig. 4 (a) with $m=0.48$ for BaFe$_{2}$As$_{2}$ is larger than the one  in  Fig. 4
(b) with $m=0.33$ for NaFeAs,  which is  also quantitatively consistent with   experiments\cite{johnstonreview}.

The remarkable character of reconstructed Fermi surfaces in
BaFe$_{2}$As$_{2}$ is the Dirac cone structures around the ($\frac{\pi}{4},0$)
and ($\frac{3\pi}{4},0$). Especially, the energy spectrum around the Dirac
cone structures in Fig. 4 (d) are similar to the experimental fitting
results\cite{richardarpes2010}. From Fig. 4 (d), we can find that the Dirac cone
structure around ($\frac{\pi}{4},0$) point is formed by a hole-like band from one
S$_{4}$ iso-spin component and a folded electron-like band from the other
S$_{4}$ iso-spin component while the Dirac cone structure around ($\frac{3\pi}%
{4},0$) point is formed by exchanging the contributions between two S$_{4}$
iso-spin components. In Fig. 5, we plotted the spectra around the ($\frac{\pi
}{4},0$) point, where the Dirac cone structures are resolved with the
red-dashed line.  The Dirac
cone structures were proposed as the universal character and robust with interaction
strength and doping\cite{ranying2009,zhout2010}. Our results clearly show  that the feature highly depends on quantitative parameter settings in a model and in general should be material-dependent.
\begin{figure}[ptb]
\begin{center}
\includegraphics[width=1.0\linewidth]{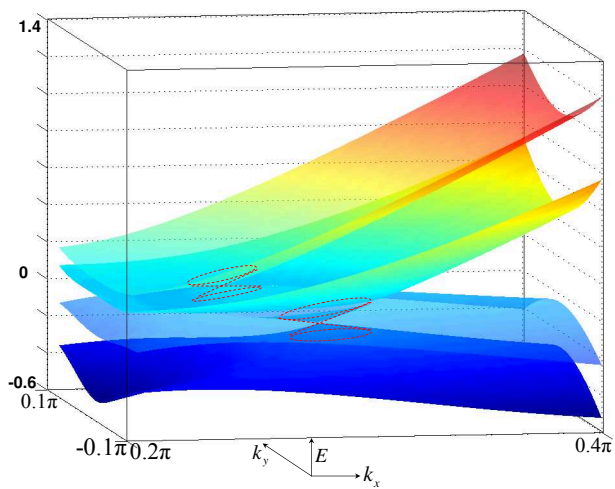}
\end{center}
\caption{(color online) The band structure around $(\frac{\pi}{4},0)$
corresponding to Fig 4. (a) (d). The Dirac cone structures are marked with the
red-dashed lines. }%
\end{figure}

\emph{Discussion and conclusion:}
Several LDA calculations have also noticed  that some properties   have to be understood in the base of the 2 Fe unit cell\cite{linch2011,colson2012}.  From above results, we clearly see this importance reflected in the starting point of  kinematics in analytic approaches.  In  most previous studies, in particular, analytic studies, calculations are performed in  models with one iron per unit cell.  In principle, the S$_4$ symmetry kinematics  is  hidden in these  models and can be revealed by performing a unitary transformation. However, when interactions are concerned and exact solutions  are impossible, the search of meanfield order parameters highly depend on the base set in kinematics, which is the key reason why an explicit S$_4$ construction becomes crucial.   The result also suggests that many physical quantities in a symmetry broken state  require   new investigation.

It is also important to note that our calculation is performed in a tetragonal lattice. It is known that a lattice distortion or a tetragonal to orthorhombic lattice transition exists at a temperature higher than or equal to the magnetic transition temperature\cite{johnstonreview}. In general, the S$_4$ symmetry is broken  in the orthorhombic lattice.  Our calculation does not include this aspect since  the model is purely electronic and  does not include an electron-lattice coupling.   The lattice distortion is likely driven by pure electronic nematism\cite{Fang2008nematic, xu2008, Eremin2011a,nematic-fernandes2010} so that  the qualitative results here will remain unchanged.  Nevertheless,  a full consideration of this aspect  is needed  in future.

In summary,  we show that the S$_{4}$ microscopic
effective model characterized by a doubling of iron unit cell explains the puzzled band reconstruction in  the magnetic state of iron-pnictides. The results provide a strong support to the kinematics described by the S$_4$ symmetry.

\emph{Acknowledgement:}  JP thanks H. Ding, D.L. Feng and T. Xiang  for useful discussion.  The work is supported  by the Ministry of Science and Technology of China 973 program(2012CB821400) and NSFC-1190024.

%
%
%
%
%
%
%
%
%
%
%
%
%
%
%

\end{document}